# Gate-tunable black phosphorus spin valve with nanosecond spin lifetimes

Ahmet Avsar[1,2], Jun Y. Tan[1,2], Marcin Kurpas[3], Martin Gmitra[3], Kenji Watanabe[4], Takashi Taniguchi[4], Jaroslav Fabian[3], Barbaros Özyilmaz[1,2]

[1] Centre for Advanced 2D Materials, National University of Singapore, 117542, Singapore

[2] Department of Physics, National University of Singapore, 117542, Singapore

[3] Institute for Theoretical Physics, University of Regensburg, Regensburg, 93040, Germany

[4] National Institute for Materials Science, 1-1 Namiki, Tsukuba 305-0044, Japan

**Two-dimensional materials offer new opportunities for both fundamental science and technological applications, by exploiting the electron's spin[1]. While graphene is very promising for spin communication due to its extraordinary electron mobility, the lack of a band gap restricts its prospects for semiconducting spin devices such as spin diodes and bipolar spin transistors[2]. The recent emergence of 2D semiconductors could help overcome this basic challenge. In this letter we report the first important step towards making 2D semiconductor spin devices. We have fabricated a spin valve based on ultra-thin (~ 5 nm) semiconducting black phosphorus (bP), and established fundamental spin properties of this spin channel material which supports all electrical spin injection, transport, precession and detection up to room temperature (RT). Inserting a few layers of boron nitride between the ferromagnetic electrodes and bP alleviates the notorious conductivity mismatch problem and allows efficient electrical spin injection into an n-type bP. In the non-local spin valve geometry we measure Hanle spin precession and observe spin relaxation times as high as 4 ns, with spin relaxation lengths exceeding 6 μm. Our experimental results are in a very good agreement with first-principles calculations and demonstrate that Elliott-Yafet spin relaxation mechanism is dominant. We also demonstrate that spin transport in ultra-thin bP depends strongly on the charge carrier concentration, and can be manipulated by the electric field effect.**

Electron spin is an important degree of freedom which can complement or even replace charge in information storage and logic devices[3]. For spin-based electronics, it is essential to have materials with long spin relaxation times at RT[1]. With respect to the material selection, semiconductors in particular offer new opportunities that are unfeasible in metal-based spintronics devices. These include doping by the electric field effect and gate controlled amplification/switching actions[4]. The boom of semiconductor spintronics started with the demonstration of the electrical spin injection into GaAs by R. Fiederling et al., also later by H. Ohno et al. [5,6]. I. Appelbaum and his colleagues further fostered this by adding silicon into the spintronics materials family[7]. More recently 2D materials such as graphene have captured the interest of engineers and scientists on the grounds of their high electronic mobility and the simultaneous ability to tune their charge carrier concentrations by the electric field effect[8]. Graphene has already been investigated very extensively in the spintronics community since the first unequivocal demonstration of RT spin injection by N. Tombros et al.,[1,9]. While the first devices showed spin lifetimes of only 0.1 ns[9], the new generation of graphene devices that have surfaces within the last year show remarkable spin lifetimes of ~ 10 ns, making graphene suitable for spin communication channels[10].

Conversely, the zero-band gap nature of graphene does not allow charge and spin conductance to be fully suppressed[1]. The gate tunable suppression of spin signal is possible in bilayer graphene, but the effect only appears at low temperatures[11]. This makes graphene less suitable for spin rectification or amplification and has triggered the search for alternative 2D semiconductor materials, such as transition-metal dichalcogenides and black phosphorus (bP). The reasons are manifold. These materials naturally support the scaled down information technologies by their atomic thickness. They have band gaps on the order of 1-3 eV, which matches both Si and GaAs[12,13]. Their electronic densities can be tuned between *n* and *p* regimes (without affecting the intrinsic properties of these materials), allowing for diode and transistor actions unthinkable for graphene[12]. In addition, and unlike graphene, 2D semiconductors absorb light efficiently enough to make opto-spintronics devices[14]. Finally, and in contrast to their bulk counterparts, the 2D materials are susceptible to proximity effects and are expected to become ferromagnetic in proximity of ferromagnetic insulators or gates[15]. All of the above make a strong case for pursuing 2D semiconductor spintronics.

There have already been *optical* measurements of the spin properties of transition metal dichalcogenides[14]. Recent Kerr rotation microscopy measurements on $MoS_2$ and $WS_2$ found long spin lifetimes in these important materials. However, also here the signal completely disappears at T > 40 K. Neither has *all-electrical* spin injection and detection with two-dimensional transition metal dichalcogenides yet been reported, not even at low temperatures.

Black phosphorus is a relatively new member in the family of 2D materials and has already been shown to be an excellent material for charge based applications[13,16,17]. Its sizeable direct band-gap[16], room temperature mobilities in the order of 1,000 $cm^2/Vs$[18], and relatively weak spin-orbit coupling make it also a very promising spintronics material. Similar to silicon (Si), it is a centrosymmetric material and Dyakonov-Perel spin relaxation is expected to be inhibited[19]. Elliott-Yafet (EY) remains as primary spin scattering mechanism and hence, the recent report by L. Li et al.[18] on very high electron mobilities suggests long spin relaxation lengths as well.

In this letter, we demonstrate that the necessary conditions for 2D semiconductor spintronics, i. e. electrical spin injection, transport and detection up to RT - can all be realized in ultra-thin bP devices. By taking advantage of the recent advances in van der Waals heterostructure fabrication methods, we fabricate non-local spin valve devices by fully encapsulating bP with hexagonal boron nitride (BN) layers. The BN/bP/BN structure is formed under an inert gas environment ensuring that the surfaces of bP are never exposed to air (Fig. 1a)[20]. The details of device fabrication and crystal growth are given in Supplementary Information. The optical image of a typical device after the deposition of ferromagnetic contacts is shown in Fig. 1b. Both the bottom BN ($BN_B$) and the top BN ($BN_T$) layers play crucial roles. The bottom $BN_B$ has been utilized as a substrate to achieve high electronic mobilities in bP by minimizing substrate related scattering sources[18]. The ultra-thin top $BN_T$ layer has two purposes. First, it protects bP from degradation upon air exposure. Hence, this prevents the formation of chemisorbed species at the interfaces that can significantly limit the charge and spin injection from metal electrode[20]. It protects bP also from polymer residues formed during the device fabrication process which has been proposed to be the source for spin relaxation in graphene[11,21]. Note that such a full encapsulation is necessary to study intrinsic transport properties of bP. Equally important, this layer acts as a tunnel barrier for efficient spin injection into bP[22]. We limited the thickness of $BN_T$ to only 3 layers such that it is sufficiently thin to form only a low barrier for electron injection but thick enough to prevent the back flow of the injected spins[9].

Spin transport measurements are carried out with standard ac lock-in technique at low frequencies and at fixed current bias ($I_{SD}$) in the four-terminal non-local spin valve geometry as a function of temperature (T), back gate voltage ($V_{BG}$) and in-plane magnetic field ($B_{\parallel}$). The spin channel between injector and detector electrodes was also characterized in the local two-terminal geometry where we recorded the source-drain current as a function of the drain voltage ($V_{SD}$), $V_{BG}$ and T. In this work, we studied a total of five different junctions in three samples with thicknesses 5 nm, 8 nm and 10 nm. All devices exhibit qualitatively very similar behavior. Here, we discuss representative results obtained in three different junctions of the ~ 5 nm thick bP sample, labeled as device A, device B and device C. Separation between injector and detector electrodes in devices A, B and C are 2.7 μm, 1.5 μm and 1.4 μm, respectively. We note that since we are measuring lateral spin valves, we can only use the field effect mobility to characterize charge transport properties. In this geometry, and unlike the Hall bar geometry, precise estimate of the charge mobility is not possible. All three devices exhibit similar low temperature and room temperature field effect mobility of ~ 1500 cm$^2$/Vs and ~ 800 cm$^2$/Vs, respectively. Unless otherwise stated, the results shown are from device A.

To avoid additional etching step, we intentionally selected bP crystals with long straight edges which are ideal for non-local geometry. Here it is also important to note that bP crystals generally cleave preferably along this zigzag direction since the ideal strength for a tensile strain in the zigzag principal crystallographic direction is much higher compared to the armchair direction[23]. This is fortuitous for spintronics studies since the effective mass is higher in the zigzag direction compared to the armchair direction and hence weaker coupling between the valence and conduction bands are expected. Such weak coupling is expected to result in smaller spin-orbit coupling strengths and consequently longer spin relaxation times[18]. The crystallographic direction of the flake is confirmed by using Raman spectroscopy. We determine the angular dependence of the Raman response by rotating the sample and measuring the scattered light polarized parallel to the incident light polarization. We choose the long straight edge of crystal as reference (θ = 0°) and at each θ values, we observe typical Raman peaks around 365, 440 and 470 cm$^{-1}$ as shown in Fig. 1c, corresponding to the $A_g^1$, $B_{2g}$ and $A_g^2$ modes[24]. Following similar analysis done by Riberio et al.,[24] we confirm that the long straight edges of the crystal is indeed along the zigzag direction consistent with recent transmission electron microscopy analysis[25].

A key element of a semiconducting spin valve device is the tunneling barrier formed at the interface of the ferromagnetic electrode (FM) and the nonmagnetic material. The spin injection efficiency depends on the quality of this barrier; uniform and pin-hole free insulating barriers yield the highest spin injection efficiencies. Necessary criteria for a tunnel barrier are a parabolic $V_{SD}$ dependence and a weak temperature dependence[26]. Hence, prior to any spin transport measurements, we first characterize the charge transport properties. Figure 1d shows the $V_{SD}$ dependence of $I_{SD}$ at 2.4 K. We observe highly non-linear $I_{SD}$-$V_{SD}$ characteristics. Such dependence is in a good agreement with the classical Brinkman Dynes and Rowell model for tunnel barriers and an average barrier height of 1.67 eV is obtained by fitting the data with the model at $V_{BG}$ = 40 V where the bP channel is most conductive[27]. This value agrees well with the values extracted previously for few layers of BN[28]. Moreover we observe nearly temperature independent $I_{SD}$-$V_{SD}$ (Fig. 1d Inset)[26]. A final confirmation that the top $BN_T$ layer acts indeed as a high quality, pin-hole free tunnel barrier will be provided later by studying the scaling of the non-local resistance as a function of the channel conductance.

Interestingly, the $V_{BG}$ dependence of $I_{SD}$ at fixed $V_{SD}$ values exhibits a strong n-type behavior as shown in Fig. 1e. It is in sharp contrast to the device characteristic of the adjacent junction without the $BN_T$ layer where typical p-type behavior is observed. At present we do not yet fully understand this behavior. It has been recently predicted that BN could decrease the effective work function of Co[29]. It has been also shown theoretically that BN can donate electrons to bP[30]. Either one of these effects could explain the n-type behavior, but the details will have to be discussed elsewhere. Here it is important to note that the total device resistance in this two-terminal measurement consists of the tunneling resistance of the $BN_T$ layers at source-drain electrode and the bP channel. The resistance contribution from the bP channel decreases continuously as $V_{BG}$ increases. Hence, at large enough gate bias ($V_{BG} \geq 50$ V), the total device resistance is dominated by the gate-independent $BN_T$ resistance[20]. This allows us to extract the contact resistance of ~ 10 kΩ. Last but not least the comparison of the I($V_{BG}$) curves of regions with and without BN (Fig. 1e inset), shows that even such ultra-thin BN layers can indeed fully encapsulate the region of interest and provide hysteresis free transport characteristics at RT.

Now we turn our attention to spin transport measurements. We start our discussion with measurements performed at low temperatures, since in bP charge mobility increases strongly with decreasing temperature[13]. Here, a spin polarized charge current is applied between electrodes 1&2 and a non-local voltage is measured between electrodes 3&4 (Fig. 1a)[9,31] as a function of in plane field $B_{\parallel}$ at fixed gate voltage. As long as the channel is in the "on-state", we observe a similar behavior for all $V_{BG}$ and all T ≤ 100 K. Figure 2a shows a representative non-local spin valve signal measured at T = 100 K and $V_{BG}$ = 30 V. Additional measurements are shown in Supplementary Information. In Figure 2a, the magnetization directions of the ferromagnets are switched by sweeping the $B_{\parallel}$ along the easy axis of the ferromagnets. Depending on the relative magnetization directions of injector (2) and detector (3) electrodes, the spin accumulation changes and this leads to a nonlocal, bipolar spin signal with a change in the non-local resistance of ΔR ≈ 15 Ω. In order to confirm the origin of the spin signal and extract key spin transport parameters, we also performed conventional Hanle spin precession measurements (Fig. 2b). For such measurement, the magnetization directions of the injector and detector electrodes are first kept parallel (antiparallel) to each other by applying an in-plane magnetic field. Then this field is removed and a magnetic field sweep perpendicular to the thin film plane ($B_{\perp}$) is performed (Fig. 2b). We observe a clear spin precession signal for both parallel and antiparallel configurations. The signal decreases (increases) for the parallel (antiparallel) configuration with increasing field and can be fitted with the solution of the Bloch equation,

$$R_{NL} \sim \int_0^\infty \frac{1}{\sqrt{4\pi D_S t}} exp\left(\frac{-L^2}{4D_S t}\right) exp\left(\frac{-t}{\tau_S}\right) \cos(w_L t) dt$$

where $L \approx 2.7$ μm is the separation between the electrodes (center-to-center distance) and $w_L$ is the Larmor frequency. This gives a spin relaxation time of $\tau_S \approx 3.2 \pm 0.2$ ns, a spin diffusion constant of $D_S \approx 0.012$ m²/s, and hence, a spin relaxation length of $\lambda_S \approx 6.2$ μm. Note that $D_S$ extracted from such Hanle measurements is in good agreement with the value obtained from field effect mobility measurements[18]. These spin transport parameters are very encouraging. We first note that despite having more than one order of magnitude lower mobility than state-of the art graphene spin valves (~ 20,000 cm²/Vs, $\tau_S \approx 12$ ns )[10], $\tau_S$ in bP is order of magnitude wise comparable.

Next, we study the magnitude of the non-local signal $R_{NL}$ itself. We observe that $V_{NL}$ increases linearly with increasing $I_{INJ}$ up to 60 μA (Figure 3b). Most of our measurements are performed with much lower values $I_{INJ}$ values of ~ 5 µA, making Joule heating related effects negligible. We further note that while there is a sample to sample variation, we observe in all our devices a non-local signal of $R_{NL} \gg 1\Omega$, which in one case even reaches ~ 300 Ω (Fig. 3b Inset, see also Supplementary Information). At low temperatures these values compare favorably with what has been observed for typical graphene spin valves (~ 2.5 Ω)[10], all metallic spin valves (~ 20 mΩ)[32] and especially with bulk semiconductors (~ 20 mΩ)[33].

Next we discuss the gate bias dependence of the non-local signal. Figure 3a shows representative normalized spin precession plots measured at different gate voltages in device B. We start our discussion at large gate bias $V_{BG}$ =50 V range, where the device is unambiguously in the "on state" and a clear spin precision signal is observed. Surprisingly, as the gate voltage is decreased, we see that for 50 V < $V_{BG}$ < 20 V the spin signal actually increases. Such an inverse scaling with gate voltage and therefore, also with channel conductivity is observed in all our devices (See Supplementary Information). It was previously already reported for graphene[34] and is a direct signature of tunneling spin injection within the framework of the 1D diffusion theory of spin transport[35] (Figure 3c). As shown in Figure 3a the spin signal becomes undetectable at lower voltages ($V_{BG}$ = 10 V) (See also Supplementary Information). That this happens suddenly and even before the channel is in the "off-state" ($V_{TH}$ = 4V) may at first seem surprising. On the other hand, it is important to keep in mind that with decreasing $V_{BG}$, not only the channel resistance but also the non-local background signal itself increases rapidly. Eventually, the magnitude of the non-local spin signal becomes comparable to the noise level (See Supplementary Information). Simultaneously, also any residual conductivity mismatch problem becomes more severe with increasing channel resistance. Finally, we discuss the spin injection efficiency in our devices. For tunneling contacts, the nonlocal spin signal is given by[35]

$$\Delta R = \frac{P^2 \lambda_S}{2w\sigma} e^{-(L/\lambda_S)}$$

where, $P$ is the spin polarization of the injected charge carriers, $w$ and $\sigma$ are the width and conductivity of bP, respectively. By inserting the $\lambda_S$ extracted from spin precession measurements, we calculate $P$ to be ~ 12 % (See supplementary Information). This value is smaller than in graphene spin valves fabricated with $TiO_2$/MgO tunneling barriers (~ 30%)[34] however compares favorably with devices using either bare MgO or $Al_2O_3$ oxide barriers (~ 5%)[11] or monolayer BN barriers (~ 2%)[22].

We turn our attention to the gate bias dependence of the spin relaxation time itself. Here we focus on the gate voltage range ($V_{BG}$ > 20 V) where a clear spin signal is observed. Figure 3c shows the $V_{BG}$ dependence of $\tau_S$ for devices A, B, and C. In all three devices, we observe a $\tau_S$ in the order of nanoseconds over a large gate voltage range. Also, common to all three devices is that the $\tau_S$ peaks near $V_{BG}$ = 30 V. For instance, device C shows the longest $\tau_S$ of ~ 4 ns at $V_{BG}$ = 30 V but decreases down to ~ 0.5 ns at $V_{BG}$ = 50 V. Similarly, $\tau_S$ decreases down to ~ 1.4 ns at $V_{BG}$ = 20 V. The origin of this peculiar behavior is unclear. Now, we focus on temperature dependence of $\tau_S$ near $V_{BG}$ = 30 V by performing Hanle measurements. Unlike in graphene, $\tau_S$ in bP shows a discernible temperature dependence and hence its correlation with momentum relaxation time ($\tau_P$) could be used to identify the limiting spin dephasing mechanism. Here we determine both $\tau_S$ and $\tau_P$ directly from Hanle measurements, but we also note that the T dependence of $\tau_P$ by independent four-terminal charge transport measurements is comparable to the one extracted from Hanle measurements (See Supplementary Information). Further, we fixed $V_{BG}$ at 30 V, since the devices at this gate voltage not only exhibit the longest $\tau_S$ but also display the highest signal to noise ratios at T = 2.4 K. While this gives the best conditions to follow the spin signal to higher temperatures, we note that a similar behavior is also reproduced at the other gate biases (See Supplementary Information). Here we note that the small variation in $\tau_S$ and $\tau_P$ over $V_{BG}$ range and the uncertainty in estimating the $V_{BG}$ dependence of $\tau_P$ make difficult to estimate the main spin dephasing mechanism just based on the bias dependences of $\tau_P$ and $\tau_S$ (See Methods). Temperature dependent measurements are more time consuming but much less prone to extrinsic and interfaces related effects than the bias dependence.

Figure 4a shows the temperature dependence of both $\tau_S$ and $\tau_P$ for device A (C). In both devices, $\tau_S$ is nearly temperature independent between 2.4 K and 60 K and decreases with increasing temperature. The extracted $\tau_S$ at 2.4 K is ~ 3.8 ns (4.8 ns) in device A (C) and decreases to ~ 2.2 ns (0.9 ns) at 250 K. Remarkably, this behavior is qualitatively very similar to the temperature dependence of $\tau_P$. Also here, $\tau_P$ is nearly temperature independent up to 60 K and decreases monotonically from ~ 225 fs (165 fs) to ~ 135 fs (48 fs) with increasing temperature. The latter behavior is not unique to our devices and has been already widely reported for bP field effect transistors. Plotting $\tau_P/\tau_S$ vs T, we see that the spin relaxation time is an almost linear function of the momentum relaxation time (Fig. 4a bottom) pointing to a mainly Elliot-Yafet type spin relaxation time over the entire temperature range. Such a behavior has so far only been reported in metals; first by M. Johnson & R. H. Silsbee in aluminum[36] which was later also confirmed by F. J. Jedema et al. [31] In this mechanism at low temperatures electrons can flip their spins due to the spin-orbit interaction coming from the host lattice, and impurity scattering providing momentum scattering. At higher temperatures phonons start to dominate the spin relaxation mechanism. Our results are the first clean demonstration of the EY mechanism in a 2D material.

To shed more light on the $\tau_P$ vs $\tau_S$ result, we have performed systematic first-principles calculations of the spin relaxation rates which would be expected from the Elliott-Yafet mechanism [37,38] for bP; details are discussed in Supplementary Information. In particular, we have calculated the spin-mixing probability $b^2$, which gives the probability to find, say, the electron with a spin down if the state has the average spin pointing up. The spin mixing probability is non-zero due to spin-orbit coupling. Since in the presence of space inversion symmetry the Bloch states are doubly spin degenerate, the calculation of $b^2$ involves making linear combinations of the states to diagonalize the Pauli spin matrix representing the chosen spin direction. With this, the spin relaxation time can be calculated using

$$1/\tau_S \approx \alpha \, b^2/\tau_P,$$

where $\alpha$ is a prefactor of the order 1, being estimated up to 4 [39], depending on the details of the impurity or phonon scattering. The momentum relaxation time is taken from the experiment. We stress that the spin-orbit coupling involved is that of the host lattice (bP) only, and not of the scatterers. The presumed light scatterers (and phonons as well) yield momentum randomization only, which is necessary for spin relaxation.

The calculated magnitudes of $b^2$ are about $4 \times 10^{-5}$ for the electron spins oriented along zigzag and armchair directions (See Supplementary Information). The extracted experimental value, using $b^2 \approx \tau_P/\tau_S$ (taking $\alpha \approx 1$) from the temperature-dependence in the Fig. 4a, is about $6 \times 10^{-5}$. Considering the uncertainty in $\alpha$, the agreement between experiment and theory is excellent. Our theory also predicts that the spin relaxation rate is highly anisotropic, expected to be 50% higher if the injected spin is perpendicular to the phosphorene planes. For holes, the spin relaxation rate should be almost twice that of electrons, with the anisotropy reaching 100% (See Supplementary Information). Similar anisotropies were predicted for anisotropic metals[40]. Note that since the spin in our Hanle experiment precesses in the phosphorene plane, we take the average values of $b^2$ for the zigzag and armchair directions (although the in-plane anisotropy is very weak (See supplementary Information)).

Finally, we study spin transport at room temperature. Note that in all our devices, the (spin) signal to noise ratio increases above 250 K. This makes the observation of a clear spin signal challenging at high gate voltages more challenging. However, since the spin signal increases as the device conductivity decreases (Fig. 3c), we can improve the signal at RT by measuring at slightly lower $V_{BG} = 20$ V instead. Furthermore, the use of a thick BN layer as a substrate makes the switching of the magnetization of the ferromagnetic electrodes at RT less "clean" (See methods). Nevertheless we observe a clear spin switching signal. The spin signal becomes much cleaner in Hanle precession measurements. In order to minimize magnetization reversal by domain wall propagation due to steep steps at $BN_T/bP/BN_T$ stack edge (Figure 1b), we introduce the following measurement protocol: we first apply initially a large positive in-plane field (red lines) before returning to zero prior to the actual Hanle precession measurements. We repeat the same measurement for large negative in-plane field (black lines) before returning to zero in-plane field. In both cases, we observe a very similar precession curves (Fig. 4-b) and hence, can extract reliably a $\tau_S$ of ~ 0.7 ns. It is important to note that electronic spin transport in bulk semiconducting systems at high temperatures is very rare. For instance, spin transport in GaAs-based spin valves is observable only up to 70 K[41]. Similarly, spin transport in Si-based spin valves in a non-local geometry was observable only at temperatures below 150K[33]. Recently, Suziki et al., have demonstrated spin transport at RT by improving the surface purification that lead higher spin injection efficiency[42]. However, the magnitude of the non-local signal in our study at room temperature is more than four-orders of magnitude higher than what has been measured in Si (~ 1 mΩ)[42]. Even compared with the highest spin signal obtained in graphene spin valves[34], our result is order of magnitude comparable. This makes ultra-thin black phosphorus a promising semiconductor material for spintronics studies and applications.

In summary, we demonstrate for the first time all electrical spin injection, transport, precession and detection up to RT in ultrathin gate-tunable bP. We show in particular that $\tau_S$ can be as high as 4 ns with $\lambda_S$ exceeding 6 µm below T < 100 K. Even at RT, we observe values of $\tau_S$ ~ 0.7 ns and $\lambda_S$ = 2.5 µm in these first generation devices. The ratio of the $\tau_S$ and the $\tau_P$ agree with our DFT based analysis which involves only the spin-orbit coupling of the host lattice, shows that Elliott-Yafet is the dominant scattering mechanism. This is important because it shows that the spin lifetimes will grow with the expected increase of the bP crystal quality, eventually limited by electron-electron scattering.

In future experiments such long spin lifetimes in conjunction with the intrinsic origin of the spin relaxation mechanism will allow to identify unambiguously the impact of bP's anisotropic crystal structure on spin transport. If confirmed, the latter has the potentially to provide directional control of spin transport[16]. Furthermore, these properties make bP also an exciting alternative to graphene to study proximity effects in highly interacting 2D van der Waals heterostructures. In contrast to graphene, effects such as induced magnetism (bP on insulating ferromagnets) or induced spin orbit coupling (bP on TMDCs or TIs) can now in principle be studied without the ambiguity arising from the importance of adatoms[43]. From a technology point of view, its semiconducting nature would finally make 2D materials also attractive for basic device concepts such spin diodes or spin transistors, since bP can be engineered to be both n type or p type[2]. Even more advanced device concepts such as gate tunable spin-orbit transfer torque devices or even device concepts depending on 2D semiconducting magnetism[44] or the interplay between the spin degree of freedom and the valley degree of freedom[45] are a step closer to being realized.

**Methods**

Methods and any associated references are available in the online version of the paper.

**Acknowledgments**

We thank S. Natarajan and Y. Yeo for their help. B. Ö. would like to acknowledge support by the National Research Foundation, Prime Minister's Office, the Singapore under its Competitive Research Programme (CRP Award No. NRF-CRP9-2011-3), the Singapore National Research Foundation Fellowship award (RF2008-07), and the SMF-NUS Research Horizons Award 2009-Phase II. M. K. acknowledges support from the DFG SPP 1538 and National Science Centre (NCN) grant DEC-2013/11/B/ST3/00824, and M. G. and J. F. from DFG SFB 689 and GRK 1570, and J. F. by the EU Seventh Framework Programme under Grant~Agreement~No.~604391~Graphene~Flagship.

**Author Contributions**

A.A. and B.Ö. designed the experiments. A.A., J.Y.T. fabricated the samples. A.A. performed transport measurements. K.W. and T.T. grew the hBN and bP crystals. M. K., M.G. and J.F. provided the theoretical work. All authors discussed the results and wrote the manuscript.

**Additional Information**

Supplementary Information is available in the online version of the paper.

Reprints and permissions information is available at www.nature.com/reprints. The authors declare no competing financial interests. Readers are welcome to comment on the online version of the paper. Correspondence and requests for materials should be addressed to B. Ö (barbaros@nus.edu.sg).

**Methods.**

**Device fabrication.** Our device fabrication starts with the transfer of ultra-thin Black Phosphorus (bP) crystal on ~20 nm hexagonal boron nitride (BN) crystal by following the method developed by X. Cui et al.,[1]. This bP/BN stack is encapsulated with another few layers of BN crystal. All the transfer process has been completed under inert gas environment with ensuring that the surface of bP has never exposed to air[2]. The final stack is annealed at 250 C for 6 hours under high vacuum conditions to remove the bubbles formed during the transfer processes. This results in cleaner interfaces between 2D layers and hence improves the bonding of bP with BN layers. We note that the devices without involving this annealing process show very low conductivity, most likely due to the low charge injection efficiency. The standard electron beam lithography technique is employed to create the electrode masks which are purposely aligned along the armchair direction of the bP flake. This process is followed by the forming of Co/Ti (30nm/5nm) electrodes under ultra-high vacuum conditions (~ $5\times10^{-9}$ Torr). Deposition rate for both Co and Ti layers is ~ 0.5 A/sec. The width of contact electrodes were varied from 400 nm to 1,000 nm in order to ensure different coercive fields. To minimize additional processing steps, we choose a single metal deposition step to form ferromagnetic electrodes. To preserve the shape anisotropy, we introduce a number of sharp kinks in the electrode shape (See Figure 1b), which prevents the magnetization reversal due to domain wall motion. However in bP due to the additional height when compared to graphene, there are now sections in the device where the ferromagnet is vertical. This design works well at low temperature but makes the switching noisier at RT.

**Growth.** BN single crystals were obtained by using temperature gradient method under high pressure and high temperature. Typical growth condition was 3 GPa and 1500 ℃ for 120 hrs. Solvent of Ba-BN system was used to obtain high purity crystals[3]. Source of BN crystals were heat treated so as to reduce oxygen impurity at 2000 ℃ under nitrogen atmosphere. The recovered BN crystals were treated by strong acid of hot aqua regia to remove residual solvent and washed by diluted water for supplying to exfoliation process.
bP crystals were obtained by melt growth process under high pressure. Typical growth condition was 2 GPa and 1200 C with slow cooling of 1℃/min. Starting material of bP crystal (5N) was encapsulated in BN capsule under Ar atmosphere. The recovered bP crystals were mechanically removed from BN capsule and then used for the exfoliation process. We note that similar high quality charge and spin transport are also achieved with the crystals purchased from HQ Graphene company, The Netherlands.

**Calculations.** The electronic structure properties of black phosphorus were calculated within density functional theory using modern state-of-the art codes. The crystal structure parameters were taken from X-ray difractometry[4]. The atomic positions were further relaxed in all directions using the quasi–Newton variable–cell scheme as implemented in the Quantum Espresso package [5] assuming for the force the convergence threshold of $10^{-4}$ Ry/a.u. with the total energy convergence condition of $10^{-5}$ Ry/a.u.; 217 $k$-points were used in the sampling of the irreducible Brillouin zone wedge. The norm conserving pseudopotential with the PBEsol exchange-correlation functional [6] was employed. For the kinetic energy cutoff we used 70 Ry and 540 Ry for the wavefunction and charge density, respectively.

The electronic structure calculations on the relaxed structure, including spin-orbit coupling, were then performed using the full-potential linearized augmented plane-wave method as implemented in the Wien2k package [7]. For the muffin-tin radius of phosphorus atoms we take 2.09 bohr and the $k_{max}$ parameter of 4 bohr $^{-1}$. For the valence electrons of the phosphorus we consider two $3s$ and three $3p$ electrons. Spin-orbit coupling has been included in the second-variational procedure in which the scalar-relativistic wavefunctions were calculated in energy window up to 5 Ry. To cure for standard DFT deficiency in the description of the band gaps in semiconductors we used the modified Becke–Johnson potential [8] with a parametrization allowing to tune the bandgap to the experimental value of 0.338 eV. In the self-consistent calculation the charge density was converged for the Monkhorst-Pack $k$-point grid with 280 $k$-points in the irreducible Brillouin zone wedge.

**Comparison of spin and momentum relaxation times.** Several factors in our device architecture might affect $V_{BG}$ dependence of spin and charge transport. These prevent us from making a precise correlation between spin and momentum relation times as a function of $V_{BG}$ alone. As discussed in the main text, the conductivity mismatch issue in our devices at low $V_{BG}$ values can dramatically affect particularly spin injection efficiency. This prevents correlating the spin and momentum relaxation times to study the spin dephasing mechanism. Moreover, our devices show surprisingly n-type behavior due to the tunneling BN barrier compared to the typical p-type bP device. Such doping from contacts might cause additional p-n interfaces which will make difficult to determine the $V_{BG}$ dependence of the momentum relaxation time accurately. Further, we would like to mention that the four-terminal, long stripe device is not as ideal as Hall bar to extract the mobility accurately. For example, marked discrepancies were reported between the Hall mobilities and field effect mobility due to the underestimated values for the gate capacitances [9,10]. One way to overcome this challenge for the non-local spin valve geometry is to compare the temperature dependence of $\tau_P$ and $\tau_S$ at *fixed* $V_{BG}$. Note that both parameters show discernible temperature dependence. In Figure 4a, $\tau_P$ at a fixed $V_{BG} = 30$ V is directly determined from the same Hanle measurements by using the relation $\tau_P = 2D_S / V_F^2$ where $V_F$ is the Fermi velocity calculated from first-principles for bP (See Supplementary Information). Note also that both temperature dependence and the magnitude of the extracted momentum relaxation times are consistent with previous reports and our additional devices extracting directly from the conductivity data (See Supplementary Information)[9,10].

**Figure Captions.**

**Figure 1│Device fabrication and charge transport characterization. a** Schematics of the device. Electrodes 5 and 6 have direct contact to bP crystal. The red spheres with arrows represent the spin diffusion during a non-local spin valve measurement. Inset shows the schematics of heterostructure. **b** Optical image of the device after the metallization process. Thickness of the BN substrate, bP channel and the encapsulating BN layer is ~20 nm, ~ 5 nm and ~ 0.8 nm, respectively. **c** Raman spectra of bP crystal. Black and red lines represent the measurements taken along armchair and zigzag directions, respectively. Note that while the intensities of $A_g^1$ and $A_g^2$ peaks are higher at zigzag direction compared to armchair direction, the intensities of $B_{2g}$ peaks are the same. This is consistent with the ref. *17*. Insets show the θ dependence of $Ag^2$ mode and the optical image of the heterostructure. The black square indicates the Raman spectroscopy location. **d** $V_{SD}$ dependence of $I_{SD}$ at fixed $V_{BG}$ values in the encapsulated device at 2.4 K. The dashed line is the fit to the BDR model. Inset shows the $V_{SD}$ dependence of $I_{SD}$ at $V_{BG}$ = 30 V. **e** Back gate voltage ($V_{BG}$) dependence of bias current ($I_{SD}$) at fixed bias voltage ($V_{SD}$) of 0.5 V and 0.1 V for the encapsulated and nonencapsulated devices, respectively. Labels in the plots correspond to the electrode numbers shown in the schematics at Fig. 1a. Inset shows the $V_{BG}$ dependence of $I_{SD}$ at room temperature. Black and red circles represent the measurements for encapsulated and nonencapsulated devices, respectively.

**Figure 2│Electronic spin transport and Hanle spin precession measurements. a** Non-local signal as a function of in-plane magnetic field. Black and red horizontal arrows represent the magnetic field sweeping directions. Vertical arrows represent the relative magnetization directions of the injector and detector electrodes. Slightly different nonlocal resistance signal observed at opposite sweeps is sample dependent and it could be due to the different switching of the additional domains in ferromagnetic contacts present at the sides of the bP flake. **b** Non-local signal as a function of perpendicular magnetic field. Measurements are performed with an injected current of 0.5 µA at 100 K and $V_{BG}$ is fixed to 30 V. The red spheres with arrows in the schematics represent the precession of spins under externally applied magnetic field.

**Figure 3 | V$_{BG}$ dependent spin transport. a** The representative Hanle spin precession curves taken at 2.4 K in the non-local geometry for injector and detector electrodes in a parallel magnetization direction. **b** Bias dependence of spin valve signal amplitude. Measurement is taken at 1.5 K and V$_{BG}$ = 20 V where the largest non-local signal was observed. **c** V$_{BG}$ dependence of spin relaxation time, non-local resistance signal and two-terminal device conductivity at 2.4 K. The bias current in the latter charge transport measurement is 5 µA, similar to the current applied in spin transport measurements.

**Figure 4 | Temperature dependent spin transport. a** Temperature dependence of spin and momentum relaxation times at a fixed V$_{BG}$= 30 V for devices A and C. The error bars for momentum relaxation time is ~ 10% of its value and they are not added into the graph for better plot clarification for device A. The bottom graph shows the temperature dependence of the momentum and spin relaxation times ratio. **b** RT Spin precession measurements performed at V$_{BG}$ = 20 V. Red (black) dotted precession curve corresponds to the data taken for the parallel electrode configuration after application of a positive (negative) in-plane magnetic field. Black curve has been offset 35 Ω from red curve for clarity. Inset shows the spin valve measurement taken at V$_{BG}$ = 20V.

# FIGURES

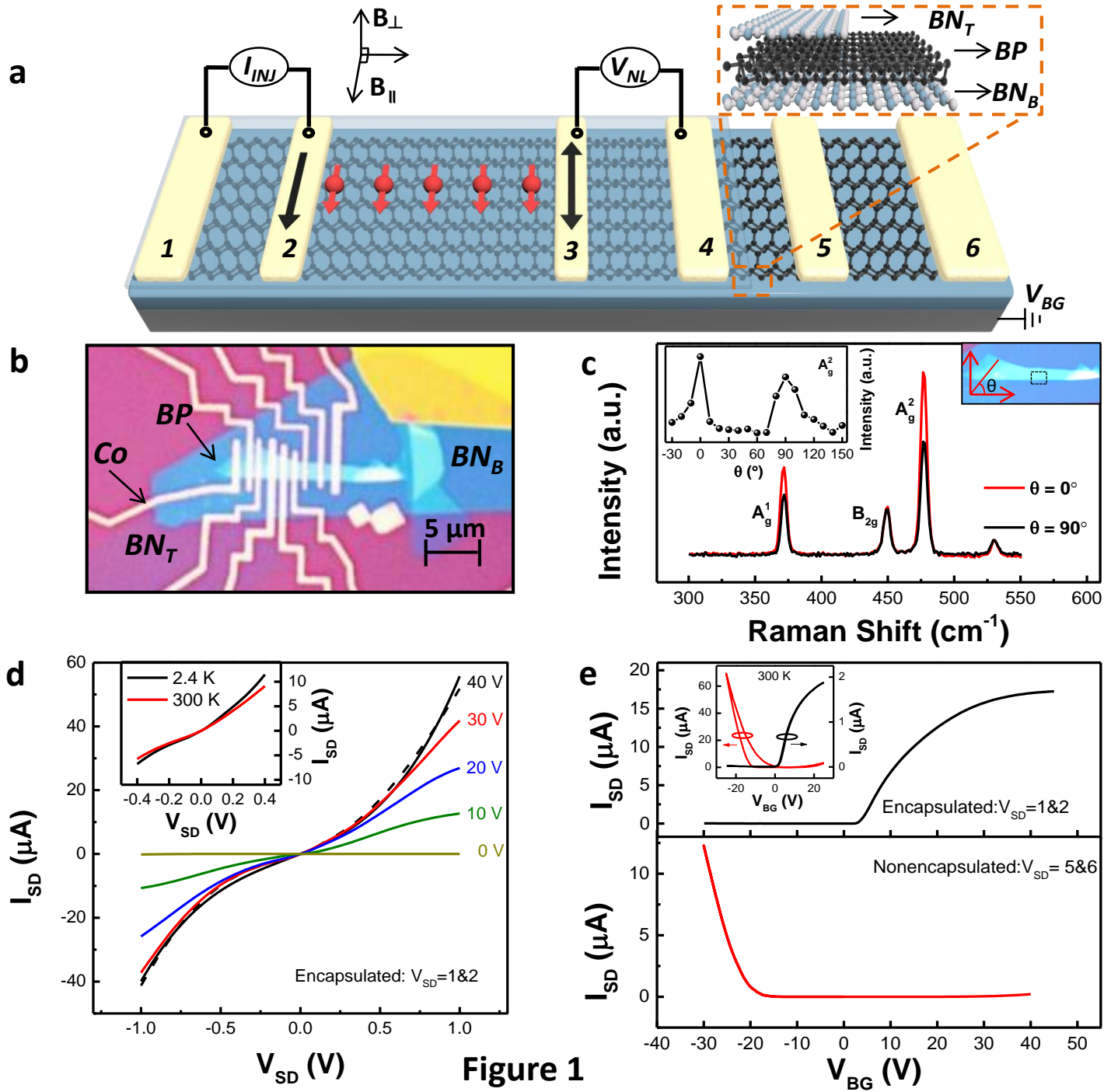

Figure 1

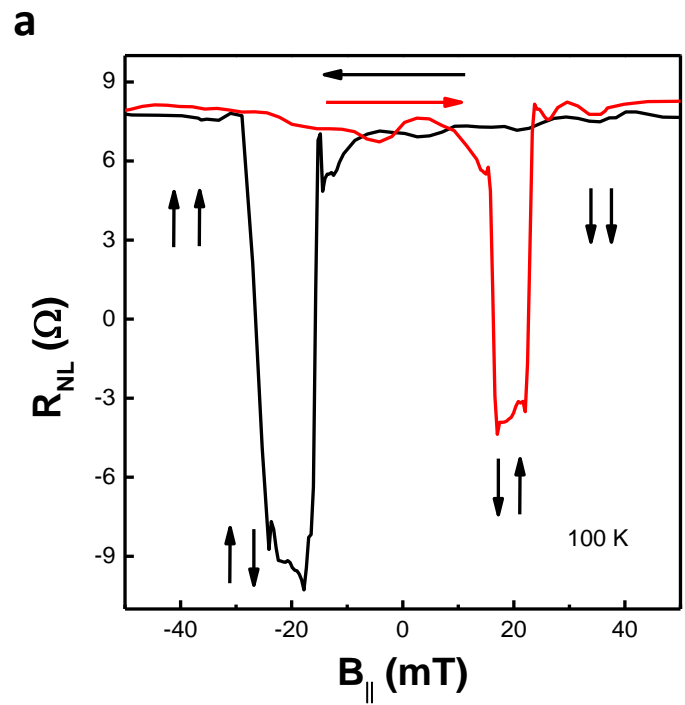 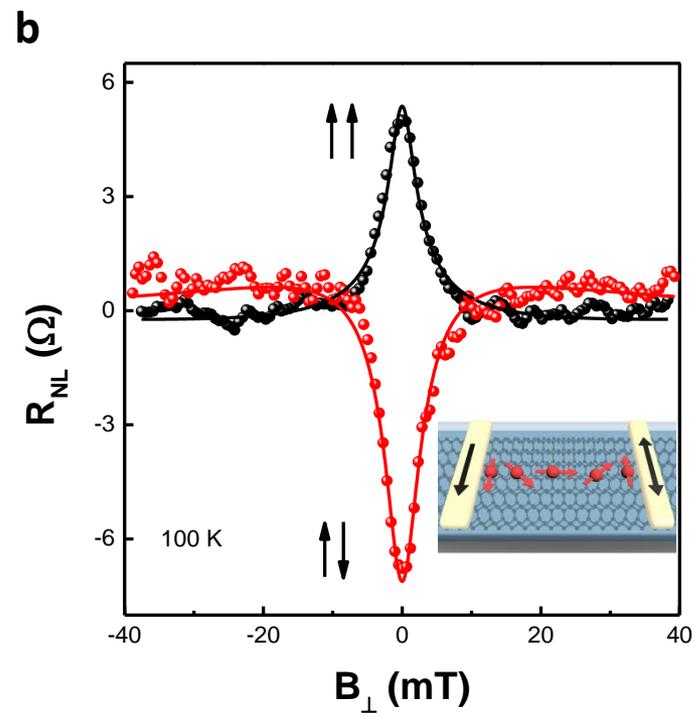

**Figure 2**

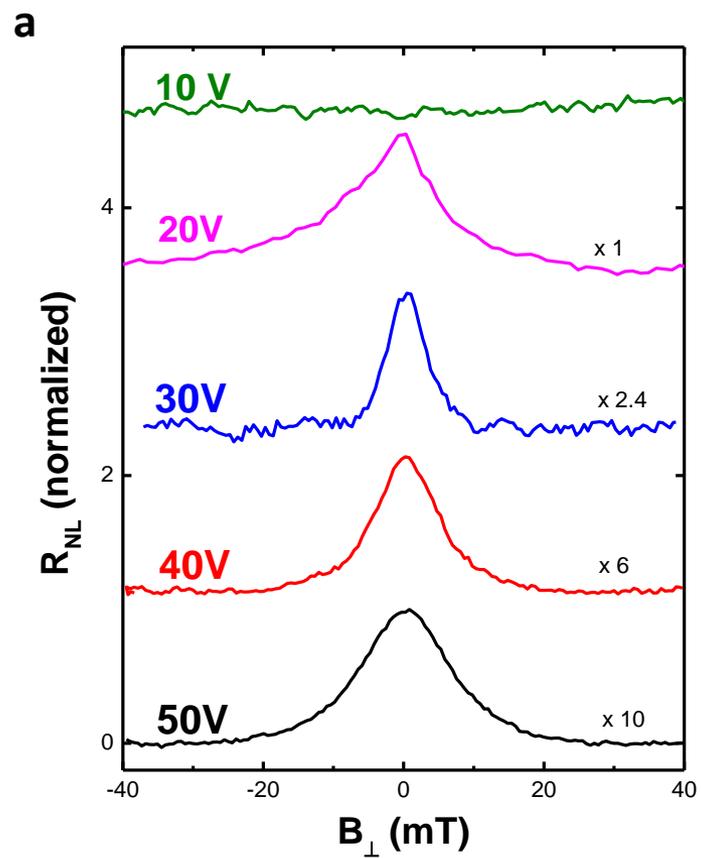
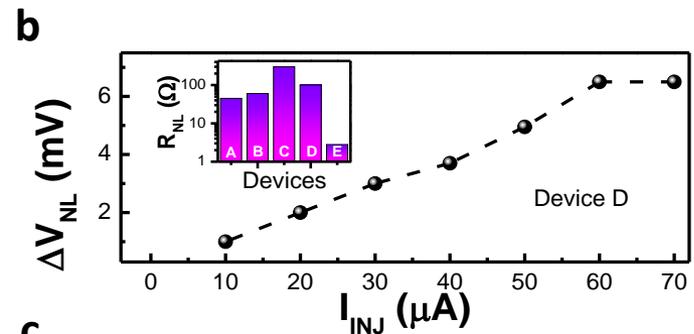
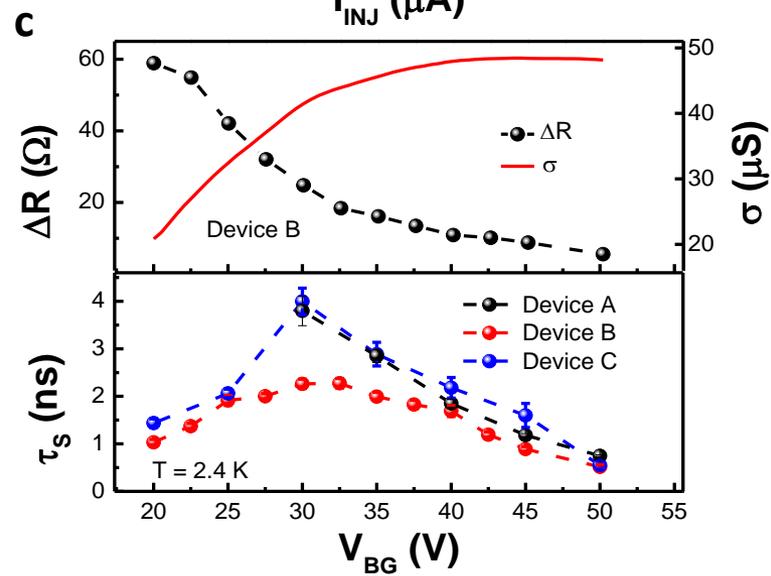

**Figure 3**

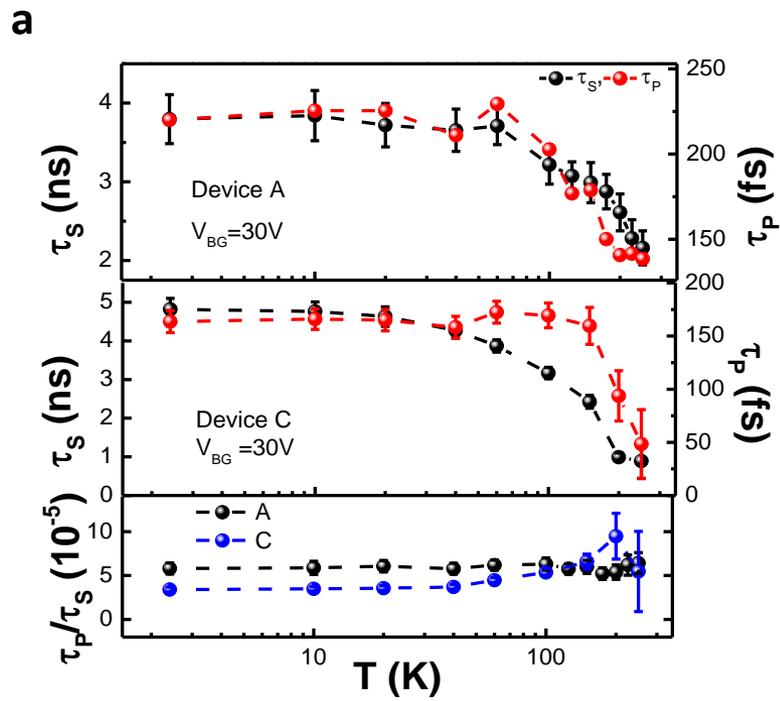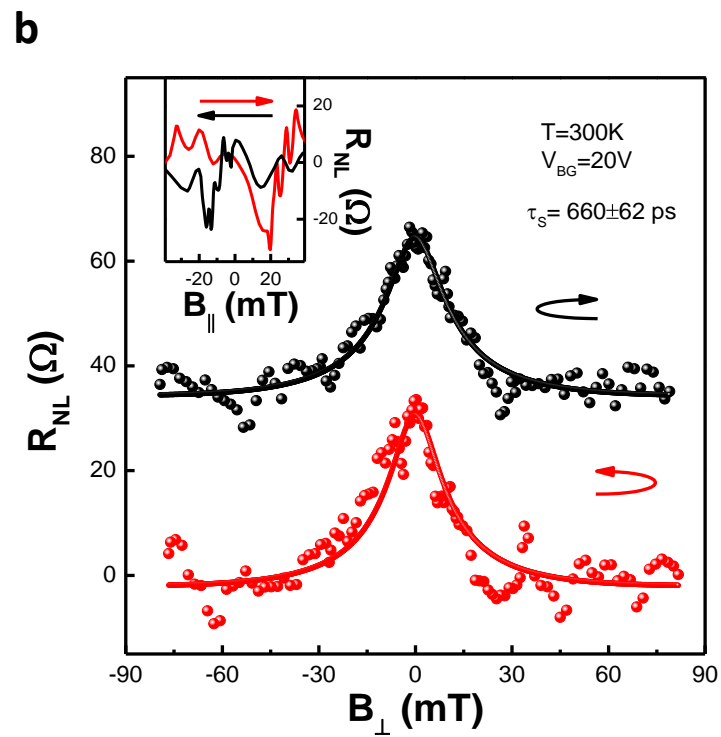

**Figure 4**